# Iterative projected gradient descent for dynamic PET kernel reconstruction

Alan Miranda[1,2] and Steven Staelens[1,2]

[1] Molecular Imaging Center Antwerp, University of Antwerp, Antwerp, Belgium

[2] µNeuro Research Centre of Excellence, University of Antwerp, Antwerp, Belgium

(Date: December 30, 2025)

**Abstract**

Dynamic positron emission tomography (PET) reconstruction often presents high noise due to the use of short duration frames to describe the kinetics of the radiotracer. Here we introduce a new method to calculate a kernel matrix to be used in the kernel reconstruction for noise reduction in dynamic PET. We first show that the kernel matrix originally calculated using a U-net neural network (DeepKernel) can be calculated more efficiently using projected gradient descent (PGDK), with several orders of magnitude faster calculation time for 3D images. Then, using the PGDK formulation, we developed an iterative method (itePGDK) to calculate the kernel matrix without the need of high quality composite priors, instead using the noisy dynamic PET image for calculation of the kernel matrix. In itePGDK, both the kernel matrix and the high quality reference image are iteratively calculated using PGDK. We performed 2D simulations and real 3D mouse whole body scans to compare itePGDK with DeepKernel and PGDK. Brain parametric maps of cerebral blood flow and non-displaceable binding potential were also calculated in 3D images. Performance in terms of bias-variance tradeoff, mean squared error, and parametric maps standard error, was similar between PGDK and DeepKernel, while itePGDK outperformed these methods in these metrics. Particularly in short duration frames, itePGDK presents less bias and less artifacts in fast kinetics organs uptake compared with DeepKernel. itePGDK eliminates the need to define composite frames in the kernel method, producing images and parametric maps with improved quality compared with deep learning methods.

## 1. Introduction

Dynamic positron emission tomography (PET) allows to measure the time dependent uptake of radiotracers. Since dynamic PET frames often have short time duration, they can present high noise, hindering quantification. To improve the image quality of dynamic PET images a wide variety of methods have been developed. For example, methods relying on temporal basis function regularization [1], regularization by fitting a kinetic model during reconstruction [2], or more recently machine learning methods [3-4], have been used to denoise dynamic PET reconstructions. The kernel reconstruction [5] is one of the algorithms used to reduce noise in dynamic PET reconstruction. In this method, spatial basis functions, calculated as a kernel matrix, are used to represent the PET image. This serves as regularization in the reconstruction, and it is especially helpful in frames with low statistics. Its calculation is relatively simple, using a kernelized version of the maximum-likelihood expectation-maximization (ML-EM) algorithm. The kernel method has been shown to outperform other, more complex dynamic PET reconstruction methods [5], including some methods using deep learning reconstruction [6].

The kernel matrix can be calculated using high quality images as priors, such as PET composite frames [5], or magnetic resonance images [7-8]. Further noise reduction can be achieved by additionally using a temporal kernel matrix with temporal basis functions [8-9].

Machine learning algorithms have been recently used to improve the kernel method. For static PET reconstruction, anatomical priors obtained from foundational medical image models have been used to construct the kernel matrix [10]. For dynamic PET reconstruction, the U-net has been used in an optimization transfer algorithm to regularize the kernel reconstruction coefficient vector [11]. In the deep kernel method, the U-net was also used to extract features from PET composite frames to then calculate the spatial kernel matrix [6].

Although machine learning used in the kernel method for dynamic PET reconstruction improves performance, these methods make use of empirically selected composite frames and therefore depend on how these composite frames are selected. If few composite frames with high quality are used, temporal information can be lost. On the other hand, using a larger number of composite frames, with shorter time duration, can preserve temporal information, but at the expense of lower image quality.

Additionally, the training calculation time in machine learning algorithms can be a limiting factor for routine use. Particularly for 3D images, computation resources and training time have to be carefully considered, often having to crop or downscale the 3D image due to limits in computation resources to train deep 3D convolutional networks [12].

Here we show that the objective function optimized in the deep kernel method using a U-net can be more efficiently optimized using projected gradient descent (PGD). For 3D images, calculation of the kernel matrix is several orders of magnitude faster using PGD compared with the deep kernel method, while both calculation methods produce images with similar image quality as demonstrated below. Moreover, building upon the PGD kernel calculation, we designed an iterative self-denoising algorithm (iterative PGD) to avoid the need to define composite frames for the kernel matrix calculation. By considering the original (fine) temporal resolution of the dynamic reconstruction, the kernel matrix can capture all the kinetic information in the scan, improving image quality in early and late time frames. Using 2D simulations and real 3D datasets we show the iterative PGD kernel outperforms all other reference methods.

## 2. Methods

### 2.1 Kernel reconstruction

The PET projection data $y_i$ ($i = 1 \dots I$ detector pais) can be modeled as the projection of the PET image $x_j$ ($j = 1 \dots J$ voxels) uisng projection matrix $P \in \mathbb{R}^{I \times J}$:

$$y = Px + r \qquad (1)$$

where $r$ are background events per detector pair. In the kernel method, the PET image is calculated by multiplying a kernel matrix $K \in \mathbb{R}^{J \times J}$ and the image coefficient vector $\alpha \in \mathbb{R}^{J \times 1}$, i.e. $x = K\alpha$. The kernelized expectation maximization (KEM) algorithm can then be used to iteratively

calculate the coefficient vector [5]:

$$\alpha^{n+1} = \frac{\alpha^n}{K^T P^T 1_I} K^T P^T \frac{y}{PK\alpha^n + r} \quad (2)$$

where $T$ indicates transpose and $1_I \in \mathbb{R}^{J \times 1}$ is a vector of ones. The image quality improvement depends on the kernel matrix and therefore how it is calculated is crucial to obtain optimal results. Originally, the use of low noise, large duration, composite PET frames $z_m$ ($m = 1 \ldots M$ composite frames) was proposed to define voxels features [5]. These features are used to calculate for each voxel $j$ the set of $N$ nearest neighbors voxels $\mathcal{N}_j$ using the Euclidian distance between features. The kernel matrix weights are finally calculated for these $\mathcal{N}_j$ neighbors using the radial basis function kernel [5]. In practice, the rows of the kernel matrix are normalized to sum 1 for improved performance [5]:

$$\bar{K} = \text{diag}^{-1}[K 1_J] K \quad (3)$$

## 2.2 Deep kernel

Recently, an alternative method to calculate the kernel matrix was proposed using the convolutional neural network U-net [6]. In this approach, the kernel matrix is modeled as a denoising matrix. The kernel matrix weights in the predefined $\mathcal{N}_j$ neighbors are calculated using the composite frames $z_m$, and a noisy version of these frames $\tilde{z}_m$, optimizing the objective function:

$$\hat{\theta} = \underset{\theta}{\text{argmin}} \sum_{m=1}^{M} \|z_m - K(\theta; Z)\tilde{z}_m\|^2 \quad (4)$$

where $\theta$ are the parameters of the U-net (Fig. 1a). The input to the U-net are the high quality composite frames $z_m$, from which optimal features are calculated. These features are used to then calculate the kernel matrix weights with the *softmax* function, which also results in the sum to 1 constraint of the kernel matrix rows. The loss of the U-net is finally calculated using (4).

## 2.3 Projected gradient descent kernel matrix

The objective function in (4) can also be formulated as a constrained linear least squares optimization (Fig. 1b), where the kernel matrix rows weights are the parameters to be optimized, constraining the solution to the non-negative orthant and to the sum-to-one hyperplane. If we arrange the $M$ composite frames in rows of matrix $Z \in \mathbb{R}^{M \times J}$, and likewise for the noisy composite frames to form $\tilde{Z} \in \mathbb{R}^{M \times J}$, the constrained least squares optimization for every kernel matrix row is:

$$\underset{\widehat{K}_{j*}}{\text{argmin}} \|z_{*j} - \tilde{z}_{*j}^N \widehat{K}_{j*}^T\|^2$$
$$\text{subject to } \widehat{K}_{j*} \geq 0 \text{ and } \sum_j \widehat{K}_{j*} = 1 \quad (5)$$

where $\widehat{K}_{j*} \in \mathbb{R}^{1 \times N}$ are the elements in the $j$-th row of matrix $K$ corresponding to the $N$ nearest $\mathcal{N}_j$ neighbors of voxel $j$, $z_{*j} \in \mathbb{R}^{M \times 1}$ is the $j$-th column of matrix $Z$, and $\tilde{z}_{*j}^N \in \mathbb{R}^{M \times N}$ is the concatenation of the columns of matrix $\tilde{Z}$ corresponding to the $N$ nearest neighbors of voxel $j$.

The equation $z_{*j} = \tilde{z}_{*j}^N \widehat{K}_{j*}^T$ is an underdetermined linear system of equations and therefore we can have multiple solutions that satisfy it. Both the non-negativity and the sum-to-one hyperplane constraints help to reduce the number of possible solutions, which serve as regularization to our problem.

Here we used the projected gradient descent (PGD) algorithm to solve (5) for every row of the kernel matrix, where at every iteration we project the solution to the non-negative orthant and to the sum-to-one hyperplane. Both projections can be analytically calculated in an efficient manner:

$$\text{Non-negative orthant projection}$$
$$x^* = \max(x, 0)$$

$$\text{Sum-to-one hyperplane projection}$$
$$x^{**} = x - \frac{\langle x, a \rangle - b}{\|a\|^2} a \quad (6)$$

where the hyperplane is defined as $H = \{x: \langle x, a \rangle = b\}$. We initialize $\widehat{K}_{j*} = 1/N$ (i.e. uniform weights with sum-to-one).

Later we show that the performance of the kernel matrix calculated with (4) or (5) is very similar, but for 3D images calculation with PGD is several orders of magnitude faster compared to using the deep kernel method.

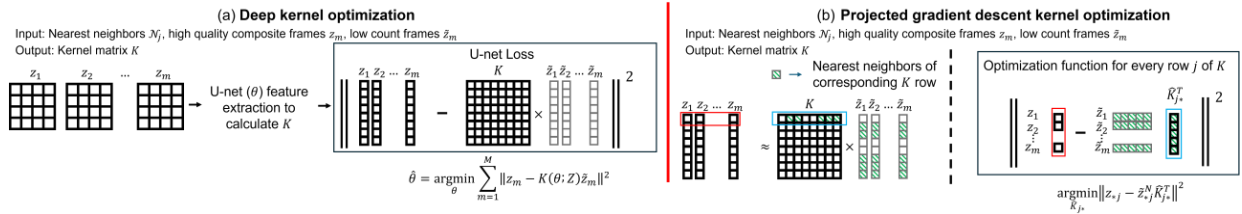

Fig. 1. (a) Deep kernel optimization for calculation of the kernel matrix using high quality $z_m$ and low quality $\tilde{z}_m$ voxel features (composite frames). (b) Projected gradient descent optimization for calculation of the kernel matrix using the same inputs as the deep kernel method. Optimization is performed independently for every row.

### 2.3 Iterative projected gradient descent kernel matrix

The use of composite frames in the kernel method has the disadvantage of failing to capture the details of structures with fast radiotracer uptake (e.g. heart, kidneys, veins). One could reduce the duration of composite frames, but this compromises noise reduction. Here we developed a method to be able to use the original dynamic reconstruction framing, i.e. with finer temporal radiotracer uptake information, for the calculation of the kernel matrix, without compromising noise reduction. The basic idea is to divide the dynamic frames is smaller groups, calculate the nearest neighbors independently for every group, and then calculate the "average" nearest neighbors. Then, similarly, the kernel matrix is independently calculated for these dynamic frames groups, using PGD, and the average of all the kernel matrices is calculated. This procedure is iteratively performed for different random combinations of frames groups, using the kernel matrix from the previous iteration to create the "high-quality" frames for the PGD optimization at the next iteration. The detailed implementation is described next (Fig. 2).

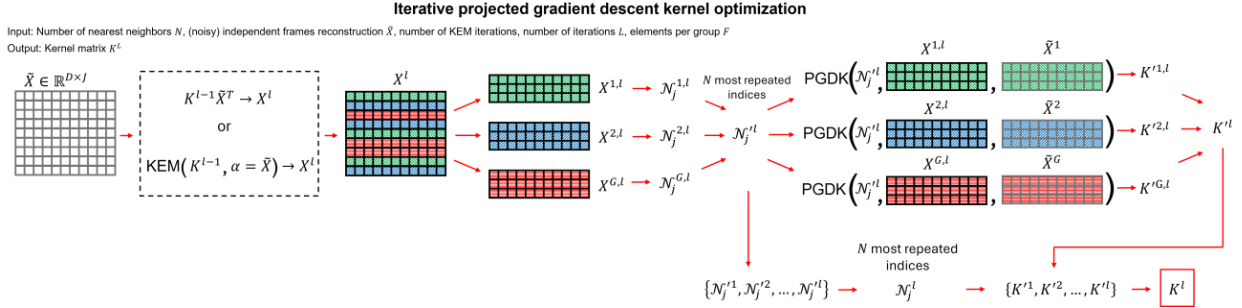

Fig. 2. Illustration of the iterative PGD kernel matrix calculation. At every iteration the dynamic frames $\tilde{X}$, with original framing, are filtered with the previous iteration kernel matrix by multiplication with the kernel matrix from the previous iteration or by running a KEM reconstruction to obtain $X^l$. The filtered frames are then divided in groups with fewer frames, and closest neighbors are calculated independently for every group. The $N$ most repeated indices are kept and used in the PGD kernel (PGDK) calculation of every group, where the noisy features are those in $\tilde{X}$ and the high quality features are those in $X^l$. The average of all kernel matrices is calculated, and considering results from all previous iterations, the most repeated neighbors indices are calculated, and their average kernel matrix weight is the kernel matrix weight for the current iteration.

*High quality reference image.* The input of the algorithm is the noisy independent frame dynamic ML-EM reconstruction with the original/fine framing scheme. The dynamic frames $\tilde{x}_j^d$ ($d = 1 \ldots D$ number of frames) are noisy voxel features, equivalent to $\tilde{z}_j$, concatenated in columns to from matrix $\tilde{X} \in \mathbb{R}^{D \times J}$. In order to create the high-quality reference version of these frames at iteration $l$ ($X^l \in \mathbb{R}^{D \times J}$) we use the kernel matrix calculated in the previous iteration ($l - 1$) to denoise $\tilde{X}$. Two different approaches can be followed to denoise $\tilde{X}$: i) Directly multiplying the kernel matrix with the noisy features, i.e. $K^{l-1}\tilde{X}^T = X^{lT}$ or ii) running the KEM algorithm with $K^{l-1}$, initializing the coefficient vector of frame $d$ as $\alpha^d = \tilde{X}_{d*}^T$, to obtain the matrix of high quality voxels features $X^l$. We tested running the algorithm only with either approach i) or ii), or a mixed approach in which we alternate i) and ii) at every iteration (see III. Results). At the first iteration we set $X^l = \tilde{X}$.

*Frames groups division.* The dynamic frames $D$ are divided in $G = ceil(D/F)$ groups with $F \geq 3$ elements per group. The $F$ frames indices of every group are calculated from a random permutation without repetitions of the $D$ frames indices. To create groups that have frames well spread in the kinetics curve, we calculate the image correlation between all pairs of corresponding $\tilde{x}^d$ image frames in the group and calculate the mean image correlation per group. From a large number of precalculated groups sets (e.g. $10^5$), we use those with the minimum correlation between the frames in every group. Therefore, at every $l$ iteration a different random set of indices is used to get $X^{g,l} \in \mathbb{R}^{F \times J}$ ($g = 1 \ldots G$ groups) matrices by concatenating the corresponding $x^{d,l} \in \mathbb{R}^{1 \times J}$ frames in these matrices' rows (Fig. 2).

*Average nearest neighbors.* Using matrix $X^{g,l}$, the $N$ nearest neighbors for every voxel $j$ are calculated independently for every $g$ group using the Euclidian distance between voxels features (columns of matrix $X^{g,l}$) to obtain the set of nearest neighbors $\{\mathcal{N}_j^{g,l}\}_{g=1}^{G}$. From this set we finally keep the $N$ most repeated elements to get the set of nearest neighbors $\mathcal{N}_j'^{l}$.

*Average kernel matrix.* Using $\mathcal{N}_j'^{l}$, we calculate the kernel matrices $K'^{g,l}$ independently for every group of frames by minimizing (5) with PGD. High quality features are those in $X^{g,l}$, while noisy features are $\tilde{X}^g$, also initializing kernel matrix row elements at the nearest neighbors as $\widehat{K}_{j*}'^{g,l} = 1/N$. The average of all groups kernel matrices is calculated to obtain $K'^{l}$.

*Final kernel matrix.* At the end of every iteration we concatenate indices $\{\mathcal{N}_j'^{l}\}$, and corresponding kernel matrix weights $\{K'^{l}\}$, effectively increasing the set size at every iteration. Again, we calculate the $N$ most repeated indices in $\{\mathcal{N}_j'^{l}\}$ to obtain $\mathcal{N}_j^{l}$, and calculate their average kernel weight by adding corresponding weights in $\{K'^{l}\}$, to get the final kernel matrix $K^l$. Since at every iteration we use a different permutation to form the group frames indices, we obtain different realizations of the kernel matrix at every iteration, which we then average with the kernel matrices from previous iterations/realizations.

Additionally, at the end of every iteration we search kernel matrix rows with large number of small value weights by counting the number of elements with a value lower than 5% of the maximum kernel row weight. If the number of elements with small weights is larger than $N/2$, we increase the nearest neighbors search window size in the next iteration for the corresponding voxel, setting a maximum window size to avoid large increase in computation time.

Fig. 2 shows a graphical illustration of the algorithm. The parameters of the algorithm are the number of nearest neighbours $N$, the number of frames per group $F$, the number of KEM iterations, and the number of iterations of the whole algorithm $L$.

## 2.4 Simulation experiments

A 2D phantom with mouse brain, liver, kidneys and heart organs was created by using real mouse TACs from a [$^{18}$F]SynVesT-1 scan, filtered with non-local means [13]. Image frames of 128×128 pixels, with size of 0.776×0.776 mm, were blurred with a 1.5 mm full-width at half maximum (FWHM) Gaussian to simulate detector spread function, and images were forward projected. Frames of 12×10 s, 3×20 s, 3×30 s, 3×60 s, 3×150 s, 9×300 s were simulated. Poisson noise was introduced to the projection data to create 10 realizations with 300 million counts (low noise) and 10 with 25 million (high noise) counts.

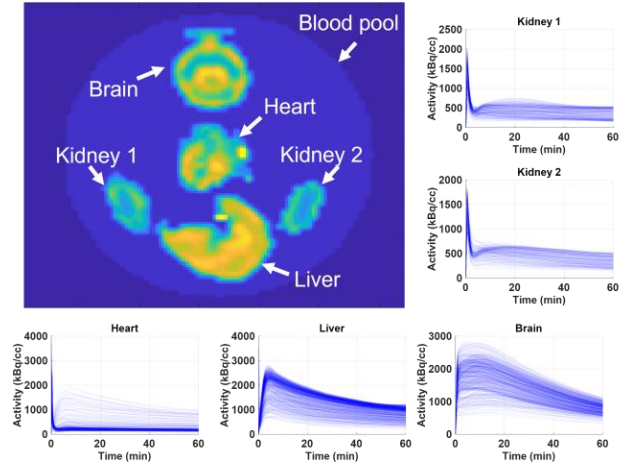

Fig. 3. 2D Simulation phantom showing the maximum pixel value in every organ, and the respective TACs of all pixels.

Bias, variance and mean squared error (MSE) are calculated for every individual organ as follows:

$$Bias^2 = \sum_{j=1}^{J}(\bar{x}_j - x_j^{true})^2 / \sum_{j=1}^{J}(x_j^{true})^2 \qquad (7)$$

$$Var = \frac{1}{N_r}\sum_{i=1}^{N_r}\sum_{j=1}^{J}(x_j^i - \bar{x}_j)^2 / \sum_{j=1}^{J}(x_j^{true})^2 \qquad (8)$$

$$MSE = Bias^2 + Var \qquad (9)$$

where $\bar{x}_j$ is the mean over all realizations, $x_j^{true}$ is the ground truth pixel value, and $N_r$ is the number of noisy realizations. Contrast was measured in the brain as the ratio of grey to white matter activity.

## 2.5 Real 3D data

Mouse scans were acquired in an Inveon microPET scanners (Siemens Medical Solutions, Inc., Knoxville, USA), and were reconstructed in an image size of 128×128×159 voxels, with 0.776×0.776×796 mm, with CT based attenuation correction.

To test the proposed methods in different noise and framing scenarios, we reconstructed the following real data whole body mouse scans. Low noise: 2 hr mouse [$^{18}$F]Fallypride scan, injected with 15.3 MBq. High noise: 90 min [$^{11}$C]raclopride scan, injected with 1.6 MBq. High time resolution: 180 s [$^{18}$F]SynVesT-1 scan, injected with 13.6 MBq. [$^{11}$C]raclopride ([$^{18}$F]Fallypride) scans were reconstructed with frames of 12×10 s, 3×20 s, 3×30 s, 3×60 s, 3×150 s, 15(21)×300 s, while [$^{18}$F]SynVesT-1with 90×2 s frames.

Kinetic modeling was performed in brain images. In the [$^{18}$F]SynVesT-1 scan, parametric maps of cerebral blood flow (CBF) were calculated using the adiabatic approximation to the tissue homogeneity model [14], using the basis function approach. An image derived input function (IDIF) measured in the left ventricle of the mouse heart was used as blood input function. For [$^{18}$F]Fallypride and [$^{11}$C]raclopride scans, parametric $BP_{ND}$ maps were calculated using the simplified reference tissue model 2 [15], with cerebellum as reference region. In all cases the standard error (SE) of the kinetic model fit is also reported.

## 2.6 Implementation details

The reference reconstruction methods are ML-EM independent frame dynamic reconstruction (MLEM), and KEM reconstruction using the kernel matrix calculated as originally proposed (Kernel [5]), the deep kernel (DeepKernel [6]), the proposed PGD kernel (PGDK), and the iterative PGD kernel (itePGDK). All kernel methods were implemented with $N = 100$. In 2D simulations, the closest neighbours search window was unrestricted for DeepKernel and PGDK, as originally defined [6]. The exact same nearest neighbours indices were used for DeepKernel and PGDK kernel matrix calculation. For Kernel and itePGDK the nearest neighbour search window was 11×11 pixels. For 3D reconstructions in all kernel methods the search window was 11×11×11 voxels. For itePGDK, with adaptive search windows size (see Sectio II.D), the maximum search window size was 15×15(×15).

For the original kernel method, a threshold of 0.8 was set on the kernel matrix weights to improve contrast [5]. The U-net for the DeepKernel calculation was implemented as described in [6], using Pytorch 2.7.1 with CUDA parallelization, training with a learning rate of $10^{-3}$, and 300 iterations. For the original Kernel, DeepKernel, and PGDK methods, composite frames of 10 min were used in 2D simulations, while for 3D data 6 composite frames with uniform time (total scan time divided by 6) were used. For the DeepKernel and PGDK, noisy composite frames were reconstructed with a subsampling factor of 10.

Projected gradient descent was implemented with CPU parallelization. To reduce calculation time, the PGD calculation was run only for kernel matrix rows corresponding to non-background voxels, setting $K_{jj} = 1$ for background voxels. We select non-background voxels by thresholding the mean of the dynamic image, setting the threshold at 0.1 the Otsu threshold [16]. We use Nesterov's acceleration with adaptive restart in the PGD calculation [17-18], setting convergence as when parameters relative change between iterations is below 0.01 %.

The KEM reconstruction step in the itePGDK method is run initializing $\alpha$ with the ML-EM image, i.e. close to the optimal maximum likelihood solution. For faster computation we also tested denoising $\tilde{X}$ by simple multiplication by the kernel matrix (see Sectio II.D). Below the effect of the different parameters in the itePGDK calculation is shown.

Reconstructions with all methods were calculated with image space resolution modeling [19-20]. To accelerate convergence in all reconstructions we used ordered subsets in the ML-EM and KEM reconstructions. For 2D simulations we used 10 subsets and 30 iterations, while for 3D reconstructions we used 16 subsets with 16 iterations for [$^{18}$F]SynVesT-1, and 32 iterations for [$^{11}$C]raclopride and [$^{18}$F]Fallypride scans. All kernel matrix calculations and reconstructions were run on a 16 core CPU (AMD Ryzen 9 5950X) with a NVIDIA GeForce RTX 3060 Ti graphics card.

## 3. Results

### 3.1 Attention maps comparison

Fig. 4 shows examples of attention maps for the different kernel methods in a cerebellum and myocardium voxel in the [$^{18}$F]Fallypride scan. In cerebellum, the Kernel attention map values have small range, being all close to the maximum value. The DeepKernel greatly improves the range of values, but their distribution is spread all over the brain, and even has some non-zero values outside the brain. The range of PGDK values is also improved but their location is concentrated in the cerebellum. The attention map values for itePGDK are also concentrated in the cerebellum but the shape is more symmetric and spans a larger region than PGDK.

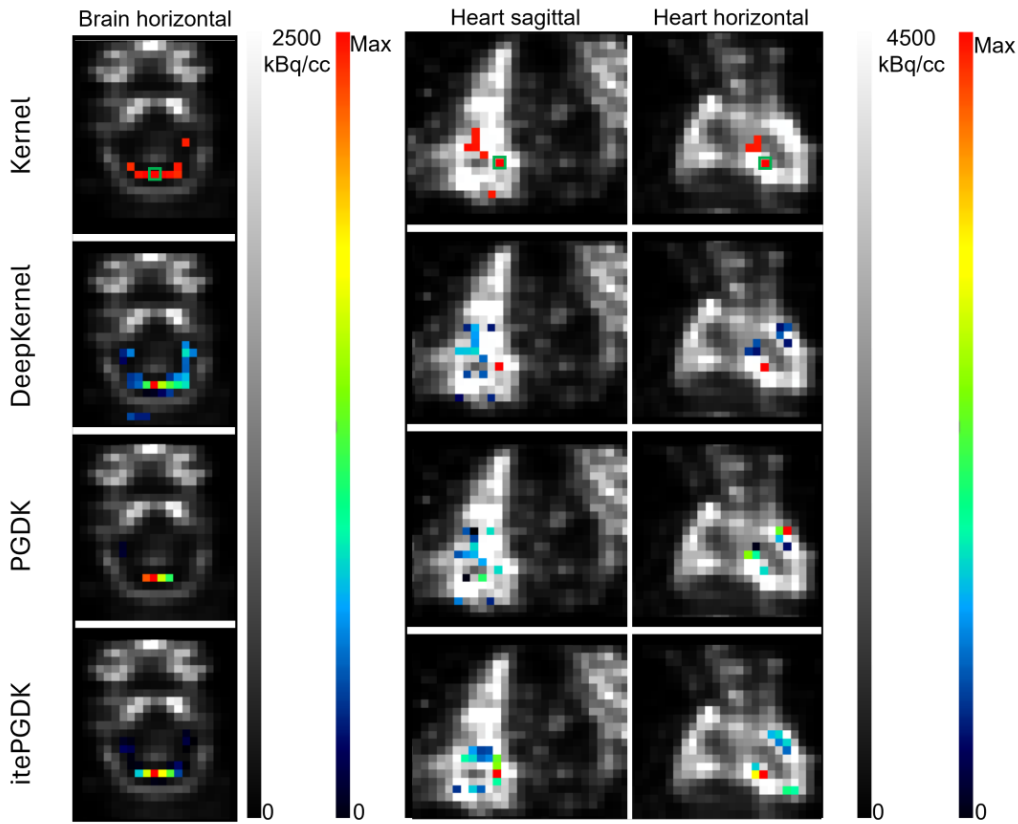

Fig. 4. Attention maps for a voxel in cerebellum and heart myocardium in the mouse [$^{18}$F]Fallypride scan. Query voxel indicated by green square in the first row. Brain horizontal plane, and heart sagittal and horizontal planes are shown. Attention map values scaled to the maximum for better visualization of the values range.

Similarly in the myocardium, the range of Kernel values is small, and is improved with the other kernel methods. Both DeepKernel and PGDK span voxels outside the myocardium (sagittal plane), while itePGK values are contained in the anatomical myocardium region, following its ellipsoid shape.

## 3.2 Effect of parameters on the itePGDK

To show the effect of group size $F$, KEM iterations, and total iterations $L$, on the itePGDK, we show the bias-variance trade-off for the 2D brain phantom, in a 10 s and 300 s frame, with different combination of the parameters (Fig. 5). Increasing the group size $F$ produce higher variance in the 10 s frame. In the 300s frames the bias is slightly reduced at larger $F$. Large number of KEM iterations reduce bias at the expense of variance increase. KEM ite = 0 corresponds to the case where the high quality reference is obtained by direct multiplication with the kernel matrix ($K^{l-1}\tilde{X}^T = X^{lT}$). Increasing the number of full iterations $L$ reduce the variance at the expense of a slight bias change (increased bias in 300 s frame). $L$ mix = 10+10 corresponds to the case where we alternate between calculation of the high quality reference by direct multiplication with the kernel matrix (10 iterations) or by running the KEM algorithm (10 iterations). With $L$ mix, the lowest variance is obtained but also with larger bias.

In summary, change in group size $F$ produce the largest difference in performance, while number of KEM iterations, and full iterations $L$ mainly change the variance, with some influence in the bias. Prioritizing bias reduction, in the following we used the itePGDK algorithm with $F = 4$, KEM ite = 150, and $L = 10$.

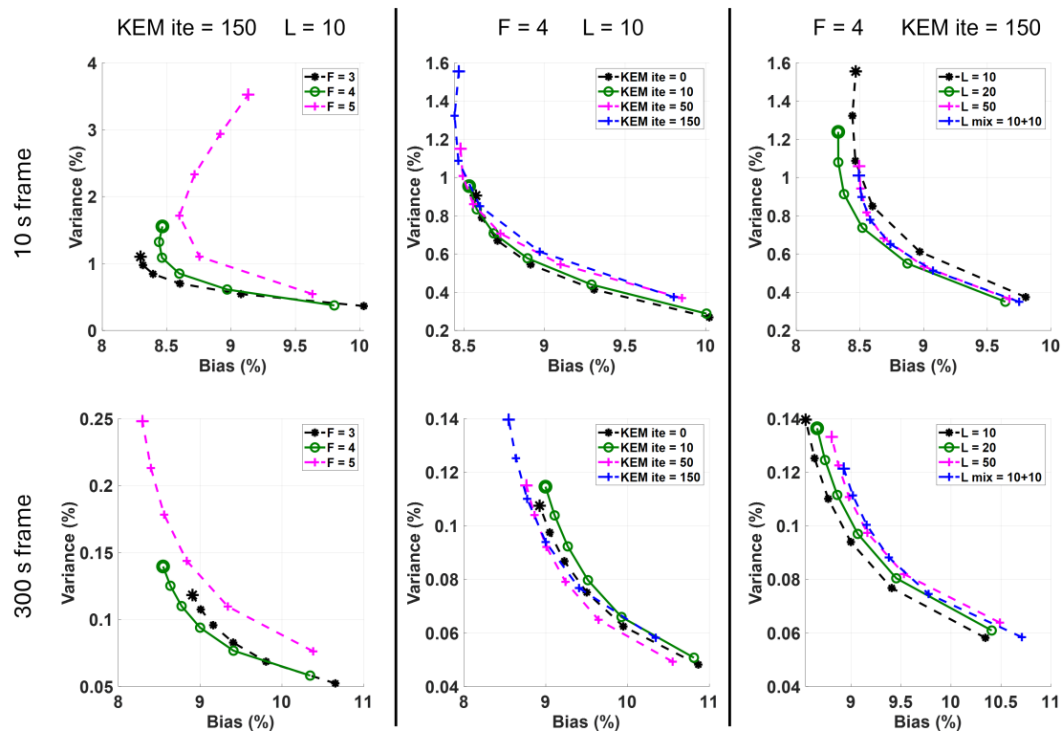

Fig. 5. Effect of the different parameters in the performance of the itePGDK method. First row: 10 s frame, second row: 300 s frame. First column: varying group size $F$, with KEM iterations = 150, and full iterations $L = 10$, second column: varying number of KEM iterations, with group size $F = 4$, and full iterations $L = 10$, third column: varying full iterations $L$, with group size $F = 4$, and KEM iterations = 150.

## 3.3 2D phantom simulation

Fig. 6 shows the bias vs variance plot at increasing iterations for the different organs in the low noise simulation. In the 10 s frame for all organs, except brain, DeepKernel and PGDK show a large bias and variance, while the lowest variance is obtained using itePGDK, and for liver and brain it also produces the lowest bias among all methos. In the 300 s frames, performance of DeepKernel and PGDK greatly improves, showing better bias and variance than MLEM and Kernel. Variance is further reduced in all organs using itePGDK, also producing the lowest bias in liver, kidney 2, and brain.

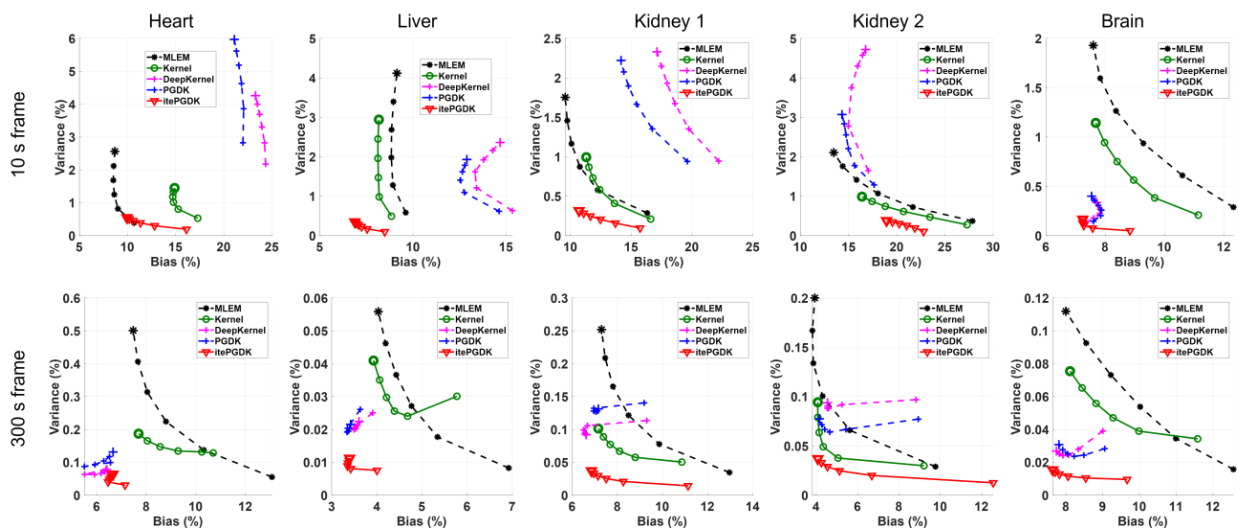

Fig. 6. Bias variance trade-off for all reconstructions, from 50 (5 iterations × 10 subsets) to 300 iterations, in 50 iterations intervals, for the different organs in the 2D phantom simulation with low noise. Top row are results for a 10 s frame, while bottom row shows results for a 300 s frame. Last iteration shown with largest symbol size.

Fig. 7 shows the average MSE in 10 and 300 s frames, and contrast vs background noise for low and high noise simulations. In low noise simulations Kernel improves MSE compared with MLEM, but in high noise simulations Kernel minimally improves MSE compared with MLEM. In low and high noise simulations, itePGDK produces the lowest MSE in all organs in 10 s frames, with similar performance between DeepKernel and PGDK. In 300 s frames with low noise, DeepKernel produces the lowest MSE in the heart, while itePGDK shows the lowest MSE in all other organs. In 300 s frames with high noise, itePGDK shows the lowest MSE, with similar performance between DeepKernel and itePGDK in kidney 2 and brain.

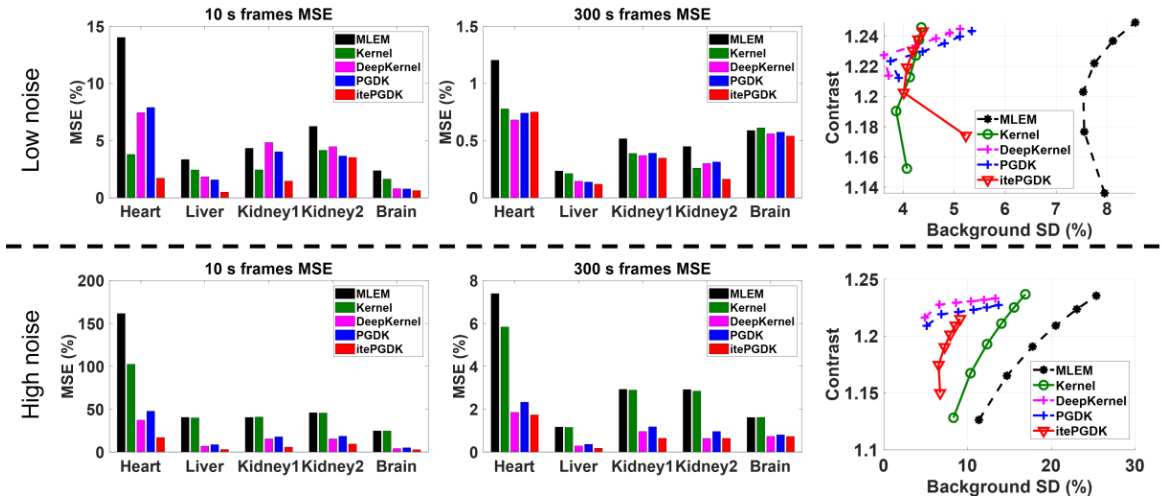

Fig. 7. Average MSE among 10 s and 300s frames (12 and 9 frames, respectively), and contrast vs background standard deviation (SD), for low noise (top row), and high noise (bottom row) simulations, with all reconstruction methods.

Contrast in low noise simulations is similar among all reconstruction methods, but lowest background noise is obtained with Kernel and itePGDK. In high noise simulations, lowest background noise is obtained with itePGDK, but also producing lower contrast.

### 3.4. Real 3D data

Fig. 8 shows the reconstruction of 2 s frames for [$^{18}$F]SynVesT-1 brain and heart, and 10 s frames for [$^{18}$F]Fallypride heart and kidney, using all methods. In 2 s frames, high noise in the MLEM reconstruction precludes visualization of structures in the brain. Kernel improves noise, with Deepkernel, PGDK, and itePGDK further reducing noise. High uptake in cortex and thalamus is observed in low noise images for example. In both [$^{18}$F]SynVesT-1 and [$^{18}$F]Fallypride heart images the shape of myocardium has less artifacts in itePGDK reconstructions compared to all other methods, showing a uniform uptake along the anatomical ellipsoid shape region. Similarly, for [$^{18}$F]Fallypride kidney images, itePGDK show the uptake in the kidney wall with improved uniformity compared to the other methods.

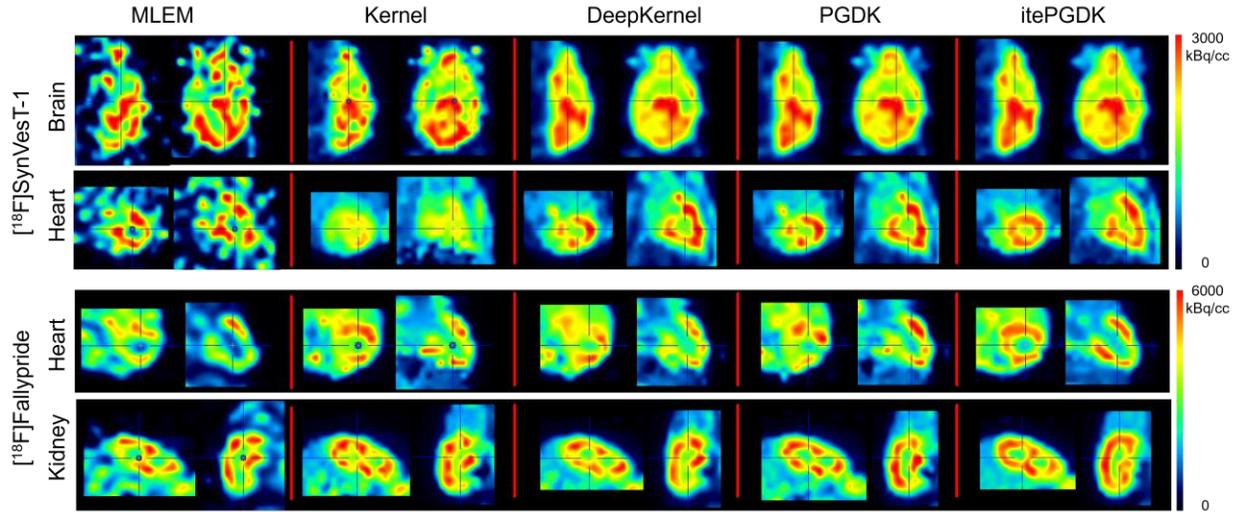

Fig. 8. Example of frames reconstructions of [$^{18}$F]SynVesT-1 (2 s frames) and [$^{18}$F]Fallypride (10 s frames) scans with all reconstruction methods. [$^{18}$F]SynVesT-1 brain sagittal and horizontal planes, and heart coronal and horizontal planes are shown next to each other. [$^{18}$F]Fallypride heart coronal and horizontal planes, and kidney sagittal and horizontal planes are shown.

Fig. 9a shows the contrast vs background (blood pool) SD in the [$^{18}$F]Fallypride scan at increasing number of iterations, calculated as the ratio of striatum to cerebellum activity. Largest background noise is observed in the MLEM reconstruction, while the lowest is obtained with itePGDK. Contrast is the largest with MLEM and PGDK, and slightly lower with itePGDK.

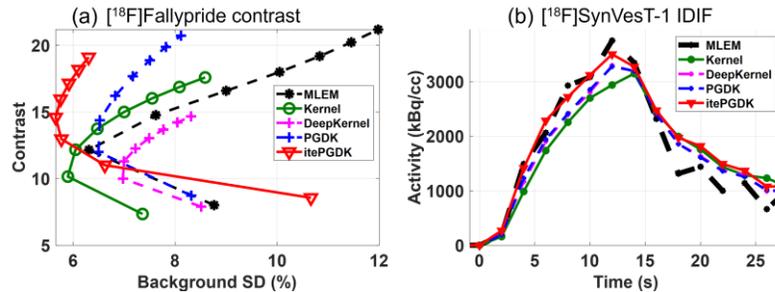

Fig. 9. (a) Background noise vs contrast measured in a 300 s frame from the [$^{18}$F]Fallypride scan using all reconstruction methods. Contrast calculated as the ratio of striatum to cerebellum activity. (b) First 25 seconds of the [$^{18}$F]SynVesT-1 image derived input function (IDIF) measured in the left ventricle of the mouse heart.

## 3.5 Brain kinetic modeling

The image derived input function used for the calculation of CBF in [$^{18}$F]SynVesT-1 images, is shown in Fig. 9b. Peak amplitude is the highest with MLEM followed by itePGDK. Rising slope is similar between MLEM and itePGDK, while the other methods present slightly slower activity increase.

Fig. 10 shows the brain parametric maps of CBF for [$^{18}$F]SynVesT-1, and $BP_{ND}$ maps for [$^{18}$F]Fallypride and [$^{11}$C]raclopride scans. MLEM CBF parametric maps present a noisy hot spot pattern and have a large fit SE. Kernel reduces the fit SE but the parametric map maintains noisy hot spots. Both DeepKernel and PGDK greatly improve fit SE and high CBF is distinguishable in thalamus and cerebellum. ItePGDK further improves fit SE and produces CBF parametric maps with smoother patterns in thalamus and cerebellum.

[$^{18}$F]Fallypride $BP_{ND}$ maps have the highest striatum contrast using MLEM and PGDK reconstructions, while DeepKernel present some loss of contrast. Fit SE is the lowest in itePGDK parametric maps. [$^{11}$C]raclopride $BP_{ND}$ maps have an asymmetric striatum $BP_{ND}$ pattern due to the high noise. DeepKernel and PGDK improve the fit SE, but the striatum $BP_{ND}$ asymmetry is still observed, also reflected in the higher left striatum SE. itePGDK improves striatum $BP_{ND}$ symmetry, also reducing the SE fit in the left striatum.

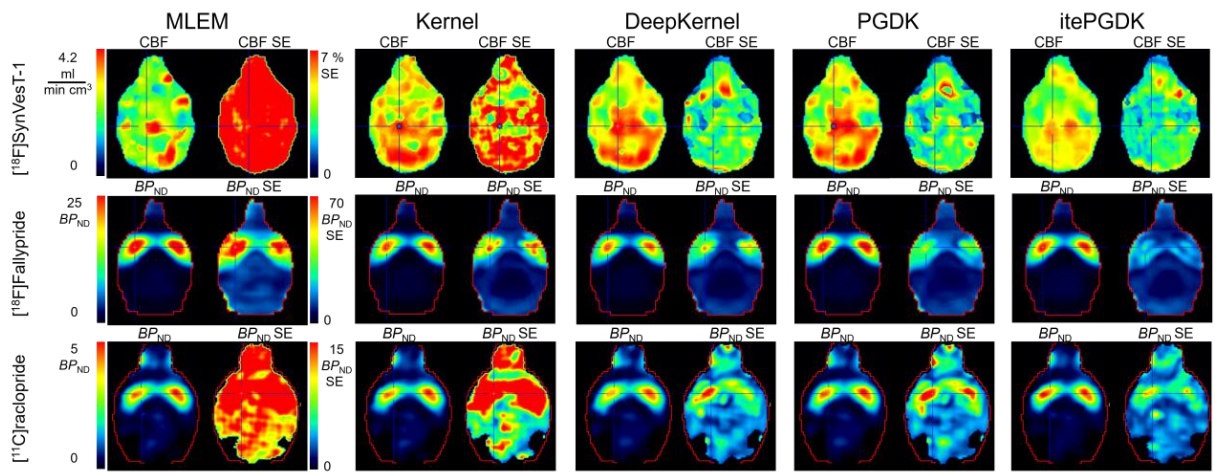

Fig. 10. First row: Brain parametric maps of cerebral blood flow (CBF) next to the standard error (SE) of the fit for [$^{18}$F]SynVesT-1. Secon row: [$^{18}$F]Fallypride $BP_{ND}$ parametric maps and fit SE. Third row: [$^{11}$C]raclopride $BP_{ND}$ parametric maps and fit SE. The same color scale shown for MLEM is applied in all images in the same row.

## 4. Discussion

Formulating the kernel matrix as a denoising matrix allows to calculate the kernel matrix weights by maximizing the similarity between high-quality and denoised noisy priors. This optimization was originally solved using the convolutional neural network U-net [6], where features are extracted from the high quality composite frames to then calculate the kernel matrix weights. Here we show that the U-net feature extraction can be bypassed by reformulating the problem as a constrained linear least squares optimization. Optimizing the kernel matrix weights which maximize the similarity between noisy and high-quality frames using PGD is much more efficient than using a convolutional neural network. Although the objective function is defined by an underdetermined linear system of equations with multiple solutions, the non-negativity constraint and confinement of solutions to the sum-to-one hyperplane allow to find meaningful solutions. Initializing the PGD solution with uniform kernel weights, i.e. the high-quality voxel feature is approximated as the average of all closest neighbors features in the noisy image, PGD refines these weights, obtaining a solution with less bias.

An advantage of the formulation of the kernel matrix as a denoising matrix is that the $\sigma$ parameter in the radial basis function kernel (RBFK) does not need to be fine tuned [5]. As observed in the Kernel attention maps, the RBFK produce kernel matrix weights with low range. In contrast, the matrix weights in DeepKernel, PGDK and itePGDK are calculated with respect to the objective function, eliminating the need of hyperparameter tuning. This produces kernel matrix weights with better range, allowing to consider the subtle differences in kinetics between the closest neighbor voxels.

A crucial aspect of the iterative PGD kernel calculation is the use of lower dimension groups of dynamic frames. Due to the so-called "curse of dimensionality" [21], using directly the noisy high-temporal resolution frames to find similarities between voxels (i.e. closest neighbors search) produce suboptimal results. It is well known that finding similarities between features deteriorates as the dimension of the features increases [21]. Sub-groups of dynamic frames are a subsampled representation of the whole dynamic sequence, which then can be used to improve the feature similarity calculation. To consider the neighbors from all groups, we calculate the most repeated indices, since noisy, or less meaningful neighbors will be less repeated among groups.

At the first iteration of the iterative PGD kernel algorithm, the noisy and high quality features are equal. Using these inputs in the PGD optimization, the trivial kernel matrix solution is the identity matrix. However, by initializing the PGD solution with uniform kernel matrix weights, a local minimum different from the trivial solution is obtained. The use of lower dimensional features (sub-groups of dynamic frames) serves to obtain a meaningful solution since calculation of linear combinations between features is improved in the lower dimensional space. Indeed, using the original high-dimensional frames (i.e. without grouping) produces a kernel matrix close to the identity matrix in the first iteration of the iterative PGD calculation.

Another step that improves performance in the iterative PGD kernel is the iterative calculation of closest neighbors using the denoised features. By using at each iteration the improved (denoised) features, the calculation of closest neighbors is also iteratively improved. In contrast, the DeepKernel and PGDK methods use a single set of closest neighbors indices for calculation of the kernel matrix.

Additionally, iterative calculation of the closest neighbors allows to adapt the search windows size in order to find better voxel neighbors. Kernel matrix weights with small value correspond to neighbors which do not significantly improve the objective function. By expanding the search window size for those voxels with large number of small kernel matrix weights, we seek to find new neighbors which hopefully contribute meaningfully to the kernel matrix. We just set a threshold on the allowed number of small kernel weight values, and a limit on the window search size to avoid excessive computation time.

The most time consuming step in the itePGDK kernel matrix calculation is running the KEM reconstruction to calculate the high image quality frames. A faster alternative is direct multiplication of the previous iteration kernel matrix with the noisy frames. This last approach has the risk of producing oversmoothed reference frames, since fidelity with the projection data is not warranted. Nevertheless, in practice, we found that simple multiplication with the kernel matrix produces good results by limiting the number of total iterations in itePGDK. In the simulation tests, 10 iterations produced good bias vs variance trade-off, comparable to the KEM denoising approach. For improved contrast, however, is recommended to use the KEM approach, which can be run for few iterations since the MLEM noisy frames are used for initialization.

Simulations show similar performance between DeepKernel and PGDK, but with slightly noisier solutions using PGDK. Since early termination of the deep kernel calculation was suggested to improve performance [6], relaxing the convergence condition of the PGDK calculation might help to produce more similar results between DeepKernel and PGDK. Nevertheless, early termination of the optimization in DeepKernel might also produce over smoothing, as observed in the 3D images contrast and in brain $BP_{ND}$ parametric maps.

Calculation of PGDK is orders of magnitude faster than DeepKernel for 3D data. Using the same hardware, calculation of kernel matrix weights for the mouse 3D scans took less than 5 seconds with PGDK, while it took approximately 9 hours with the U-net approach, i.e. more than 3 orders of magnitude faster calculation time with PGDK. In other PET denoising applications it has also been reported that deep 3D convolutional neural networks can take considerable computation time on large images [22].

Simulations show the poor performance of Kernel, DeepKernel, and PGDK in early frames due to the use composite frames. As seen in the heart TACs, the activity rapidly increases to then decay to background level. Therefore, composite frames of more than a few minutes do not capture the heart high uptake pattern. In theory, the duration of the composite frames can be fine tuned, but a balance between precise temporal information and image quality needs to be considered. The itePGDK method eliminates the need for composite frames selection by using the final intended framing of the dynamic PET image. Iteratively denoising this image then provides the high image quality reference needed for the PGDK calculation. We tested the robustness of this approach in real data noisy ([$^{11}$C]raclopride) and high-temporal resolution ([$^{18}$F]SynVesT-1) dynamic frames, obtaining improved results compared to the other methods in these cases.

Particularly in early, short duration frames, itePGDK shows improved organ uptake patterns (e.g. heart myocardium and kidneys), as also seen in the low MSE in simulations 10 s frames. Since composite frames precision in capturing slow kinetics information is better, improvement in later time frames is observed in DeepKernel and PGDK. However, also in these late time point frames itePGDK produce improved results, as reported in simulations MSE and contrast in real 3D scans.

Kinetic modelling parametric maps also show the best performance using itePGDK in terms of model standard error fit. When using an image derived input function for kinetic modelling is important to avoid over smoothing in the heart region. As observed in [$^{18}$F]SynVesT-1 and [$^{18}$F]Fallypride images, a better definition of the heart uptake is observed using itePGDK compared with DeepKernel and PGDK. This is also noticed in the [$^{18}$F]SynVesT-1 input function, in which itePGDK has a faster activity increase and higher peak magnitude compared with DeepKernel and PGDK. This could also explain the difference in CBF magnitude calculated with itePGDK in comparison with DeepKernel and PGDK.

Brain $BP_{ND}$ parametric maps show loss of contrast using DeepKernel, possibly caused by the early termination of the optimization [6] which in turn can produce oversmoothed images. PGDK on the other hand improves contrast, which could be the result of the kernel matrix optimization reaching a lower objective function value. When checking the final objective function value (4), it was indeed smaller for PGDK compared with DeepKernel.

Interestingly, the symmetry in [$^{11}$C]raclopride striatum $BP_{ND}$ pattern is improved using itePGDK. This could be due to the use of the adaptive size of the closet neighbor search window. Due to the small (left or right) striatum size, and high contrast, most close neighbors calculated in the initial search window have kernel weights close to zero as calculated with PGDK. This is identified by the itePGDK algorithm, and therefore the search window is expanded in the next iterations, allowing to detect better voxel neighbors in the other side of the striatum. The striatum uptake symmetry is therefore improved by considering information from the opposite side.

The itePGDK method can be used to perform PET scans with lower injected activity to for example reduce patient radiation dose, or to allow lower injected mass in small animal studies to better satisfy tracer concentrations [23]. As shown with the high-temporal resolution reconstruction example, it can also be used to perform these type of reconstructions in scanners with lower sensitivity compared to whole body PET scanners. These high-temporal resolution reconstructions then allow to use newly developed kinetic models (e.g. CBF modelling [14]).

## 5. Conclusions

Using the more efficient calculation of the kernel matrix with projected gradient descent, we developed an iterative algorithm which does not require high image quality priors. Both the high quality images and the kernel matrix are iteratively calculated, using lower dimensional representations of the PET kinetics data. This allows to capture the kinetics information of fast uptake organs with less bias, while maintaining noise reduction. Simulations and parametric imaging in real data 3D scans show the superior performance of the iterative projected gradient descent method compared to the calculation of the kernel matrix using deep learning methods.